\font\tenit=cmti10
 1
 1
 1

\documentstyle[pre,epsf,aps]{revtex}
\def\ee{\end{equation}}
\def\be{\begin{equation}}
\title{ Compact parity conserving percolation in one-dimension} 
\author{\sl N\'ora Menyh\'{a}rd\\ {\tenit  Research Institute for Solid State
Physics,
 H-1525 Budapest,P.O.Box 49, Hungary}}
\author{\sl G\'eza \'Odor\\ {\tenit Research Institute for Technical
Physics and Materials Science,
H-1525 Budapest, P.O.Box 49, Hungary}}
\begin{document}
\maketitle
\vskip .3cm
\begin{abstract}
Compact directed percolation is known to appear at the endpoint 
of the directed percolation critical line
of the Domany-Kinzel cellular automaton in $1+1$ dimension.
Equivalently, such
transition occurs at zero temperature in a magnetic field
H, upon changing the sign of H, in the one-dimensional 
Glauber-Ising model,  with
well-known exponents characterising spin-cluster growth.
We have investigated here numerically these exponents
in the non-equilibrium generalization 
(NEKIM) of the Glauber model  in the vicinity of the  
parity-conserving phase
transition point of the kinks. Critical fluctuations on the
level of kinks are found to affect drastically
the  characteristic exponents of spreading of spins
 while the
hyperscaling relation holds in its form appropriate for compact clusters.

\end{abstract}
\bigskip
\bigskip
PACS numbers: 05.70.Ln, 05.50.+q

\section{Introduction}
In the  one-dimensional Domany-Kinzel automaton (DKCA)\cite{DoKi,Kinz}
the state $\sigma(i,t)$ of site $i$ at time $t$ depends on 
$\sigma(i-1,t-1)+\sigma(i+1,t-1)$, ($\sigma(i,t)=0,1$). Of the
conditional probabilities $p(\sigma(i-1,t-1)\sigma(i+1,t-1)|\sigma(i,t))$
the independent ones are denoted by $p_0=p(00|1)$, 
$p_1=p(01|1)=p(10|1)$ and $p_2=p(11|1)$.
All sites are updated simultaneously in the process.
The phase diagram of the DKCA in the $(p_1,p_2)$ plane, exhibits a 
line of (second order) critical points of
directed percolation universality class, which line ends at the
so called compact directed percolation point (CDP).
This point is situated on the line $p_2=1$,$p_0=0$ at $p_1=1/2$.
By crossing this point (changing the sign of $p_1-1/2$) the
transition  is a  first order one between two ordered phases 
(empty and full or, equivalently, using the spin variable 
$s(i,t)=2\sigma(i,t)-1$,
all spins up and all spins down).
 The characteristic critical exponents of the CDP transition
are  known exactly and a hyperscaling relation has  
also been derived for such transitions in d dimensions \cite{DoKi,DiTr}. 
 For the spreading process
of a single $\sigma(i,0)=1$ in the sea of zeros  the exponents $\delta_s$, 
$\eta_s$ and $z_s$
 defined at the transition point for the power-law time-dependences of
the density of $1$'s $n_s\propto t^{\eta_s}$, 
the survival probability $P_s(t)\propto
t^{-\delta_s}$ and the mean square distance of spreading
$<R_s^2(t)>\sim t^{z_s}$
have been obtained as $0$, $1/2$ and $1$, respectively \cite{DiTr}. 
(In the following subscript $s$ refers to spins for all
the quantities. Without subscript the corresponding quantity for kinks
is meant, except for $\nu$).
For the 
parallel(time-direction) and perpendicular(space-direction) 
coherence lengths $\nu_{\|}$ and $\nu_\perp$
resp., as well as for  the dynamical
 critical exponent $Z$ 
Domany and Kinzel has obtained the exact results:
$\nu_{\|}=2$, $\nu_\perp=1$, $Z=2$, respectively  (which means only two
exponents as by definition
$\nu_{\|}=Z\nu_{\bot}$). The above mentioned hyperscaling law \cite{DiTr}
\begin{equation}
\eta_s + \delta_s =dz_s/2
\label{eq:hy}
\end{equation}
is fulfilled with the above exponents. More generally,
Dickman and Tretyakov argue, that eq.(\ref{eq:hy}) is valid  at 
first order transitions and it should apply to cases,
whenever power-law growth produces compact 'colonies', developing from
 single seeds. In ref.\cite{DiTr} 'compactness' is  clearly defined:
it is meant that the density of colonies in surviving samples
remains finite for $t\to\infty$.

It is obvious, that
the above sketched  
$1+1$ dimensional CDP transition
 is equivalent to that in the 1-d (ferromagnetic) Glauber-Ising 
model \cite{gla63}
at $T=0$,because the
symmetry as well as the kinetics are the same. Changing the parameter $p_1$
of the DKCA around $p_1=1/2$ corresponds to introducing a magnetic field
$H$ into the spin-flip probability $w_i$ of the Glauber-Ising model
(Section II.) and changing its sign.

On the basis of this equivalence it is of some interest to investigate
the same spreading problem in the framework of the nonequilibrium
generalization  \cite {racz,MN94} of the kinetic Ising model (NEKIM) 
(Section III.), where, 
in some range
of its parameters, there is a continuous transition between a single-domain
and a multidomain state. The order parameter of this transition is the
density of kinks. The critical fluctuations of this so called parity
conserving (PC) transition  \cite{gra84,gra89,MN94,jen94,dani,kim94}.
have
pronounced effect on the underlying spin system,  as was found earlier
by the authors \cite{MEOD96} both in case of static and dynamic
exponents {\it in situations of quenching from $T=\infty$} (random
initial states).
These investigations will now be completed  by studying,
 via numerical simulations,
the spin spreading process  at the PC point (Section IV). It is
 found that the characteristic 
exponents differ from those of the  CDP transition, as
could be expected, but still basic
similarities remain. Thus the transition which takes place upon changing
the sign of the magnetic field is of first order and its exponents
satisfy eq.(\ref{eq:hy}). Accordingly it can be termed as 'compact',
and we will  call it 
 compact
parity conserving transition
(CPC).\footnote{In a previous paper of the present authors \cite{OM98}, 
devoted to
damage spreading investigations of different non-equilibrium
one-dimensional models, the issue  of a CPC transition has already been
raised.}
The static magnetic critical exponent $\Delta$ is
also determined at the PC point.

\section{Glauber-Ising model}

 The $d=1$ Ising model with Glauber kinetics is exactly solvable.
In this case the critical temperature is at $T=0$,
the transition is of first order. 
 We recall that 
$p_T=e^{-\frac{4J}{kT}}$ plays the role of $\frac{T-T_c}{T_c}$ in 1d 
 and in the vicinity of $T=0$ critical
 exponents can be defined as powers of $p_T$, thus e.g. that of the coherence
length, $\nu$, via  $\xi\propto {p_T}^{-\nu}$.
In the presence of a  magnetic field $H$,( when the Ising Hamiltonian
is given by ${\mathcal{H}}=-J\sum_{i}s_is_{i+1}- H\sum_{i}s_i$, $s_i=\pm1$),
the magnetization is known
exactly. At $T=0$ 
\begin{equation}
m(T=0,H)=sgn(H). 
\label{eq:sgh}
\end{equation}
Moreover, for $\xi \gg 1$ and
$H/{kT} \ll 1$ the the exact solution reduces to
\begin{equation}
m\sim 2h\xi\,; \qquad h=H/k_{B}T .
\label{eq:m1}
\end{equation}
In scaling form one writes:
\begin{equation}
m\sim \xi^{-\frac{\beta_s}{\nu}}g(h\xi^{\frac{\Delta}{\nu}})
\label{eq:m2}
\end{equation}
where $\Delta$ is the static magnetic critical exponent.
Comparison of eqs. (\ref{eq:m1}) and (\ref{eq:m2}) results in
  $\beta_s=0$ and 
$\Delta=\nu$ . These values are the ones
 well known  for the 1-d Ising model. It is clear that the 
transition is discontinuous at $H=0$ also when changing $H$ from positive
to negative values, see eq.(\ref{eq:sgh})( In the following the order
of limits will always be meant as: 1). $H\to 0$ and then 2). $T\to 0$).

The kinetics of the Ising model in a magnetic field
has been formulated by Glauber\cite{gla63}. In its most general form the 
spin-flip transition rate
for spin $s_i$ 
sitting at site $i$  is :

\begin{equation}
{w_i}^h=w_i(1-\tanh{h}s_i)\approx w_i(1-hs_i)
\end{equation}
\be
w_i = {\frac{\Gamma}{2}}(1+\tilde {\delta} s_{i-1}s_{i+1})\left(1 - 
{\gamma\over2}s_i(s_{i-1} + s_{i+1})\right)
\label{eq:wi}
\ee
where $\gamma=\tanh{{2J}/{kT}}$ ($J$ denoting the coupling constant in
the Ising Hamiltonian), $\Gamma$ and $\tilde\delta$ are further
 parameters.  This model will reach the same equilibrium state
as the Ising model in a magnetic field. 

For the case $\tilde\delta=0$, $\Gamma=1.0$, which is usually
referred to as the Glauber-Ising model, 
 $Z=2$ ($Z$ is the usual dynamic critical
exponent) is also a well-known result.
The non-equilibrium generalization of the kinetic Ising model to
be used later on will be given a short review in the next section.

\section{The NEKIM model}

In the NEKIM model, besides the spin-flip 
transition-rate eq.(\ref{eq:wi}), taken at $T=0$, 
also a nearest neighbour mixing of spins
with probability $p_{ex}$ 
is applied at each time step of
the simulation.
The 
 spin-exchange transition rate of nearest neighbour spins
(the Kawasaki\cite{kaw72} rate at $T=\infty$) is
$w_{ii+1}={1\over2}p_{ex}[1-s_is_{i+1}],$
where $p_{ex}$ is the probability of spin exchange.
 Spin-flip and spin-exchange are then applied
alternately.
The model was originally
proposed  and investigated  for values 
${\tilde\delta}\geq 0$ at finite temperatures in \cite{racz}.
It is , however, at $T=0$ and  for negative values of $\tilde\delta$, that
in this system a  second order phase transition takes place\cite{MN94} 
for the {\it  kinks} 
from an absorbing
to an active state, which belongs to the parity conserving (PC)
universality class .
The order parameter is the density of kinks,
at the PC point it decays in time as a power law $n_{kink}\propto
t^{-\alpha}$, with $\alpha=.285(3)$. 

The absorbing phase is double degenerate, an initial state decays 
algebraically to the stationary state, which is one of the absorbing
ones (all spins up or all spins down, provided the initial state 
has an even number of kinks) and the whole absorbing phase behaves
like a critical point with power law decay of correlations, like the 
Glauber-Ising point ($\tilde\delta=0$, $p_{ex}=0$). 

Now let us look at the PC transition from the point of view of the
underlying spin system.
The above mentioned first order transition
at $T=0$ of the Ising system disappears at the PC point and is, of course,
absent in the whole active phase of the kinks.
The fluctuations of this PC transition exert a pronounced effect on the 
underlying spin system as found earlier \cite{MEOD96} thus e.g.
the the classical dynamical exponent $Z$, defined, as usual through
the relaxation time $\tau_s$ of the magnetization, $\tau_s\propto\xi^Z$,  
was found to be 
$Z=1.75(1)$ instead
of the Glauber-Ising value of $Z=2$. 
In this case one approaches the PC point from the temperature 'direction',
by decreasing it to $0$ (the effect of temperature is to create
kink pairs inside of ordered spin-domains). 
On the other hand we can also look at the
transition by changing a characteristic parameter (chosen by us to
be $\tilde\delta$) of NEKIM  through the critical
point $\tilde\delta_c$ and fixing the other two. 
As a function of $\epsilon=\mid 
(\tilde\delta-\tilde\delta_c)\mid$ the
transition - on the level of spins - is again a first order one of type
order-disorder.
Namely, taking initial states with an even number if kinks, the
magnetization of the stationary state has a jump at $\epsilon=0$.
The same is true when changing the magnetic field $h$ from negative
to positive values at $\epsilon =0$. Thus for the spins
the value of the (static) critical exponent $\beta_s$ is zero,
in all the three 'directions'  of departing from
PC ($p_T$, $\epsilon$ and
$h$), as mentioned above.
(For simulational results see \cite{MEOD96,OM98}).

In the following we will choose the same PC transition point as in
previous works \cite{MN94,MEOD96},
and make simulations at and around this point by changing the magnetic
field $h$.
 The parameters chosen are:
$\Gamma=.35, p_{ex}=.3, \tilde\delta_c=-.395(2)$. 
 In these previous simulations the spin-flip part has been
applied using two-sublattice updating. After that
 we have stored the states of the spins  and made L
(L is the size of the system) random attempts of exchange using 
always the 
stored situation for the states of the spins before updating.
All these together has been counted as one time-step
of  updating. ( Usual Monte Carlo update in this last step
enhances the effect of $p_{ex}$ and leads to $\tilde\delta_c=-.362(1)$).

\section{Spin-cluster-growth simulations}

Time - dependent simulations have proven to be a very efficient method
for determining critical exponents (besides  the critical point
itself)\cite{Grato,Li,zheng96}.
On the basis of eq.(\ref{eq:m2}),
the $t$-dependence of the magnetization in scaling form can be written
as
\begin{equation}
m(t,h)\sim t^{-\frac{\beta_s}{\nu Z}}{\tilde g}(ht^{\frac{\Delta}{\nu Z}})
\label{eq:m3}
\end{equation}
Such  form can be  used in a quench from $T=\infty$ to $T_c$ and was 
exploited also in \cite{MEOD96},though at $h=0$, using temperature as
a second variable, for determining  mainly static critical exponents
of the spins at the PC point.

In the following we will further study the influence
of the PC transition on the spin system from
a different point of view. Instead of starting with an initial state
of randomly distributed up- and down-spins with zero average
magnetization as in the above mentioned simulations of quenching,
 we will now investigate 
the evolution of the
nonequilibrium system from an almost perfectly magnetized initial
state (or rather an ensemble of such states) . This state is
prepared in such a way that  a single up-spin is placed in the sea
of down-spins at $L/2$.
Using the language of kinks (or particles, in the BARW model\cite{TaTr,jen94}),
 this corresponds
to the
usual initial state of two nearest neighbour kinks placed at the
origin. 
The quantities usually measured of the forming clusters are the order-parameter
density, the survival probability $P(t)$ and the average mean square 
size
of spreading $<R^2(t)>$ from the center of the lattice. 
At the critical point these quantities
exhibit power law behaviour in the limit of long times; more generally
we can write 
\begin{equation}
n_s(t,h)\sim t^{\eta_s}g_1(ht^{\frac{\Delta}{\nu Z}})
\label{eq:ns}
\end{equation}
for the deviation of the spin density from its initial value,
$n_s=m(t,h)-m(0)$,
\begin{equation}
P_s(t,h)\sim t^{-\delta_s}g_2(ht^{\frac{\Delta}{\nu Z}})
\label{eq:P}
\end{equation}
for the survival probability and
\begin{equation}
<R_s{^2}(t,h)>\sim t^{z_s}g_3(ht^{\frac{\Delta}{\nu Z}})
\label{eq:R2}
\end{equation}
for the average mean square distance of spreading from the origin.
The argument of the scaling functions above has been taken from 
eq.(\ref{eq:m3}); but now at the PC point instead of the
Glauber-Ising one. Thus the exponents $\Delta$, $\nu$ and $Z$ in the
above equations
take values appropriate 
 at the PC point. We note here that the coherence length exponent $\nu$
appearing above is basically different from the $\nu_\bot$ and
$\nu_{\parallel}$ generally used 
in the context of DP transitions or in connection with the {\it kinks} 
in NEKIM. 
Namely, $\xi_{\bot}\propto \epsilon^{-\nu_{\bot}}$ with 
$\epsilon$ denoting the deviation from the PC point in the 'direction'
 of the quantity driving the phase transition \mbox{($(\epsilon=
 \mid{\tilde\delta-
{\tilde\delta}_c}\mid$ here).} Moreover, $\nu_{\parallel}=\nu_{\bot}Z$.
Z is independent of the abovementioned 'directions', 
characterizes solely the transition 
point, as it should be \cite{MEOD96}.

We have measured  $n_s(t)$ and  $P_s(t)$ at and in the vicinity of the critical point 
with initial configuration of a single up-spin  at the origin
in the see of down-spins and allowing the  system  to evolve 
according to the rule of NEKIM as described above. 
Averaging has been  taken over runs with different sequences of random 
numbers during the evolution.
\begin{figure}[h]
  \centerline{\epsfxsize=12cm
                   \epsfbox{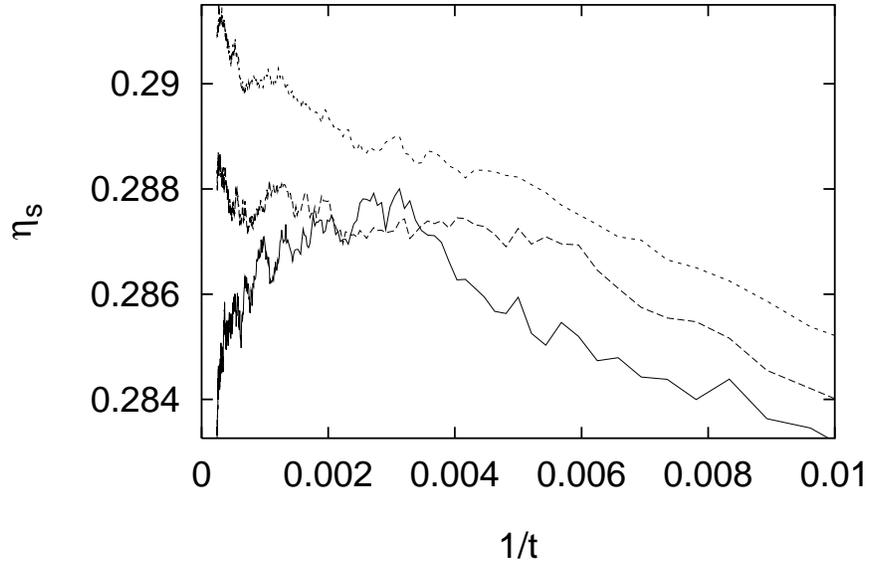}
                   \vspace*{6mm}
  }
  \caption{Local slopes of the spin density $ n_s(t)$ 
   for zero magnetic field near the PC point. $-\tilde\delta=0.393, 0.394, 0.395$
   (from bottom to top). 
   The best scaling result is $\eta_s=.288(4)$. 
   In the averaging the number of independent runs was $3-5\times 10^6$.}
   \label{fig1}
\end{figure}
\begin{figure}[h]
  \centerline{\epsfxsize=12cm
                   \epsfbox{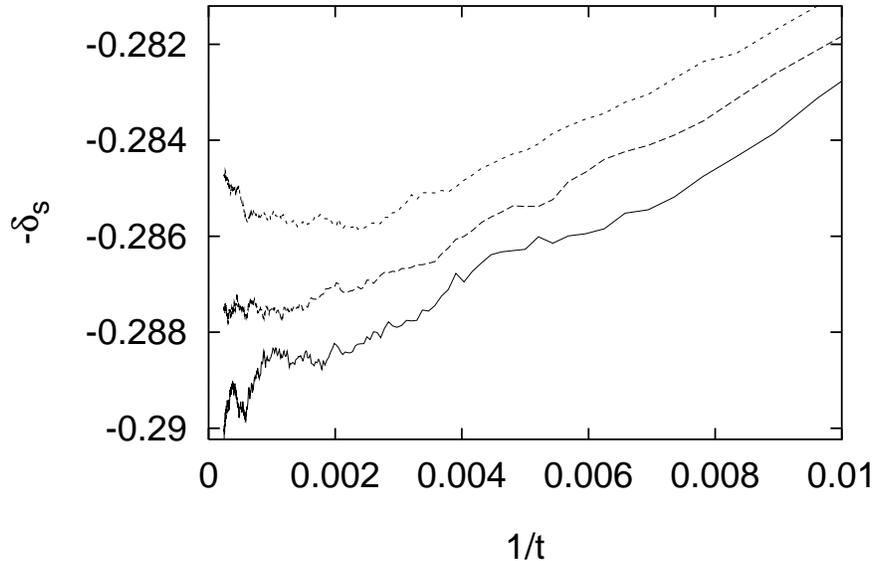}
                   \vspace*{6mm}
  }
  \caption{Local slopes of the survival probability $P_s(t)$
   for zero magnetic field in the vicinity of the PC point.
    $-\tilde\delta=0.393, 0.394, 0.395$ (from bottom to top). 
   The best scaling result is\ $\delta_s=.287(3)$. In the
   averaging the number of independent runs was the same as for Fig.\ref{fig1}
  }
  \
 \label{fig2}
\end{figure}
Fig.\ref{fig1}  and Fig.\ref{fig2} show the local slopes (for definition see 
e.g.\cite{jen94}) $\eta_s$ and $-\delta_s$,
respectively. As the survival probability must be the same for spins and kinks
(if the minority spin dies out, kinks also disappear and vice versa)
$\delta_s=\delta$. 
The same applies also for the RMS size of the cluster. 
As no result has been reported before for
$\delta$ in the NEKIM model, exhibiting Fig.\ref{fig2} has its own merits.
$\eta_s$, however, is an independent new exponent.
\begin{figure}[h]
  \centerline{\epsfysize=12cm
                   \epsfbox{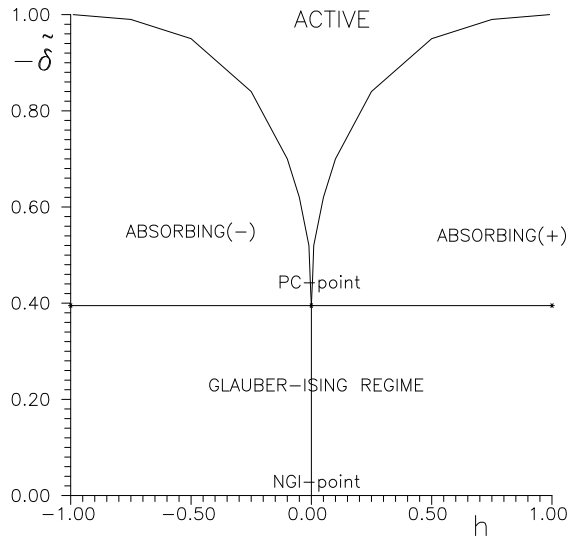}
                   \vspace*{4mm}
  }
  \caption{Phase diagram of NEKIM in the $(h,-\tilde\delta)$ plane.
The chosen PC-point is at $\tilde\delta=-.395$. 
For lower values of  $-\tilde\delta$,
in  the Glauber-Ising regime, the vertical line connecting
the PC and the NGI points (at $h=.0$) consists of all CDP points
with its characteristic critical exponents.
The simulations around the PC point have been done here for
$h>0$ in the interval $0\leq h \leq .1$. The other parameters of NEKIM 
in the whole plane are as follows: $\Gamma =.35$ and $p_{ex}=.3$.} \label{fig3}
\end{figure}
(Besides results at the PC point we have also carried out detailed
simulational studies at a point in the so called Ising phase,
namely for $\tilde\delta=0$, $\Gamma=.35$ and $p_{ex}=.3$. This point
is a non-equilibrium one due to the non-zero value of $p_{ex}$,
and is marked on Fig.\ref{fig3} with NGI(non-equilibrium Glauber-Ising)
on the abscissa. The results which we have obtained via
simulations at this point  (Fig.\ref{fig4}) are, within error, the same as 
for the (exactly solved) Glauber-Ising case.)

Fig.\ref{fig5} shows the the asymptotic values for large times of $P_s(t,h)$,
for different values of $h$ in the range of $h=.005 - .1$. For the exponent 
${\beta_s{'}}$ defined through
\begin{equation}
\lim_{t\rightarrow \infty} P_s(t,h)\propto h^{{\beta_s}'}
\end{equation}
the value $\beta_s{'}=.445(5)$ has been obtained. Fig.\ref{fig6} and 
Fig.\ref{fig7} show the scaling functions, eqs. (\ref{eq:ns}) and (\ref{eq:P}), 
respectively.
\begin{figure}[h]
  \centerline{\epsfysize=9.5cm
                   \epsfbox{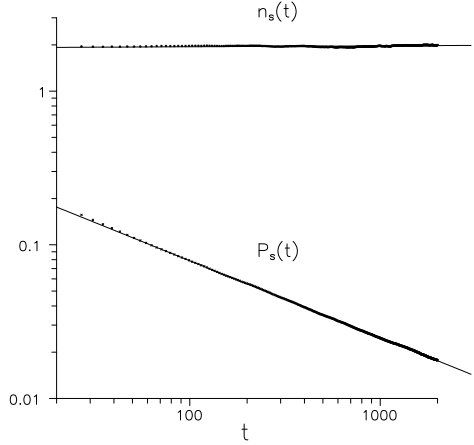}
                   \vspace*{4mm}
  }
  \caption{The scaling function $n_s(t)\propto t^{\eta_s}$ and
$P_s(t)\propto t^{-\delta_s}$ at the NGI point of Fig.\ref{fig3}
( $p_{ex}=.3$, $\tilde\delta=0$). The power  law fit of the
data shown gives ${\eta_s}^{NGI}=.0006$ and  ${\delta_s}^{NGI}=.500(5)$. 
Number of independent runs in the averaging was:$(10-60)\times 10^4$}
\label{fig4}
\end{figure}
\begin{figure}[h]
  \centerline{\epsfysize=9.5cm
                   \epsfbox{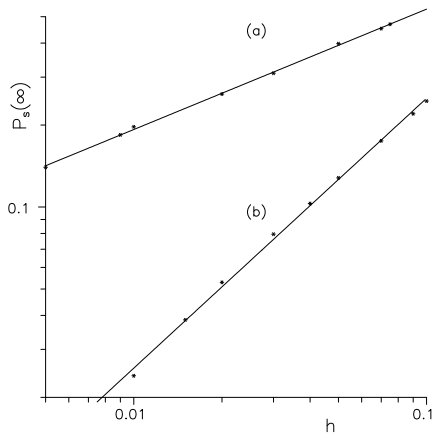}
                   \vspace*{4mm}
  }
  \caption{Level-off values of the survival probability $P_s(t,h)$ 
for large times at different values
of h. The straight line (a) is power law fit near the PC point with
 ${\beta_s}{'}=.445(5)$. Points around the straight line (b) show simulational
results of the same quantity for  the NGI point, giving ${{\beta_s}{'}}^{NGI}
=.99(2)$. Number of independent runs in the averaging was: $10^4 - 10^5$.}
\label{fig5}
\end{figure} 
\begin{figure}[h]
  \centerline{\epsfysize=9.5cm
                   \epsfbox{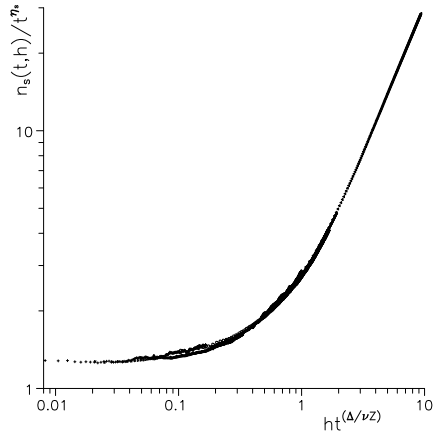}
                   \vspace*{4mm}
  }
  \caption{The scaling function $n_s(t,h)$.
 The different curves correspond to the following
values of the parameter $h$: $h=.05, .01,.009,.003,.001,.0005$.
  The values of the parameters
$\eta_s$, $\Delta$ and $\nu Z$ are given in the text. 
Number of independent runs in the averaging was: $4\times 10^4$.}
\label{fig6}
\end{figure} 
Here the best fit for the scaling-together of data with different
values of $h$ could be achieved with $\Delta=.49(1)$, 
using the measured values  
 $\delta_s=.285$, $\eta_s=.285$,
and that of $\nu Z$ from former studies, $\nu Z=.777$ \cite{MEOD96}.
\begin{figure}[h]
  \centerline{\epsfysize=9.5cm
                   \epsfbox{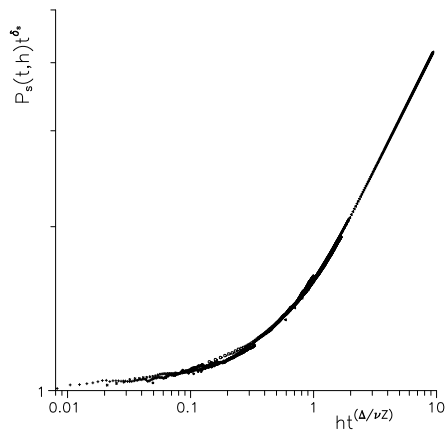}
                   \vspace*{4mm}
  }
  \caption{The scaling function $P_s(t,h)$. The values of the parameters 
$\eta_s$, $\Delta$ and $\nu Z$ used are given in the text. Number of
independent runs and values of $h$ are the same as for Fig.\ref{fig6}}
\label{fig7}
\end{figure} 
Data for different
values of $h$ scale together sufficiently well when considering the
relatively poor statistics ( averages over $4\times 10^4$ samples, typically).
The scaling law 
\begin{equation}
\beta_s{'}=\frac{\delta_s \nu Z}{\Delta}
\label{eq:scaling}
\end{equation}
following from eq.(\ref{eq:P}) is satisfied with the above values of the exponents,
within error. We note here that $\beta_s{'}$ can be connected with
$\beta_{kink}=\beta_{kink}{'}$,  using in eq(\ref{eq:scaling}) $\delta_s=
\delta$ and the definition of $\beta_{kink}{'}$ from ref.\cite{mend}
with the result: $\beta_s{'}=\beta_{kink}\nu/{(\nu_{\bot}\Delta)}$.

\begin{figure}[h]
  \centerline{\epsfysize=11cm
                   \epsfbox{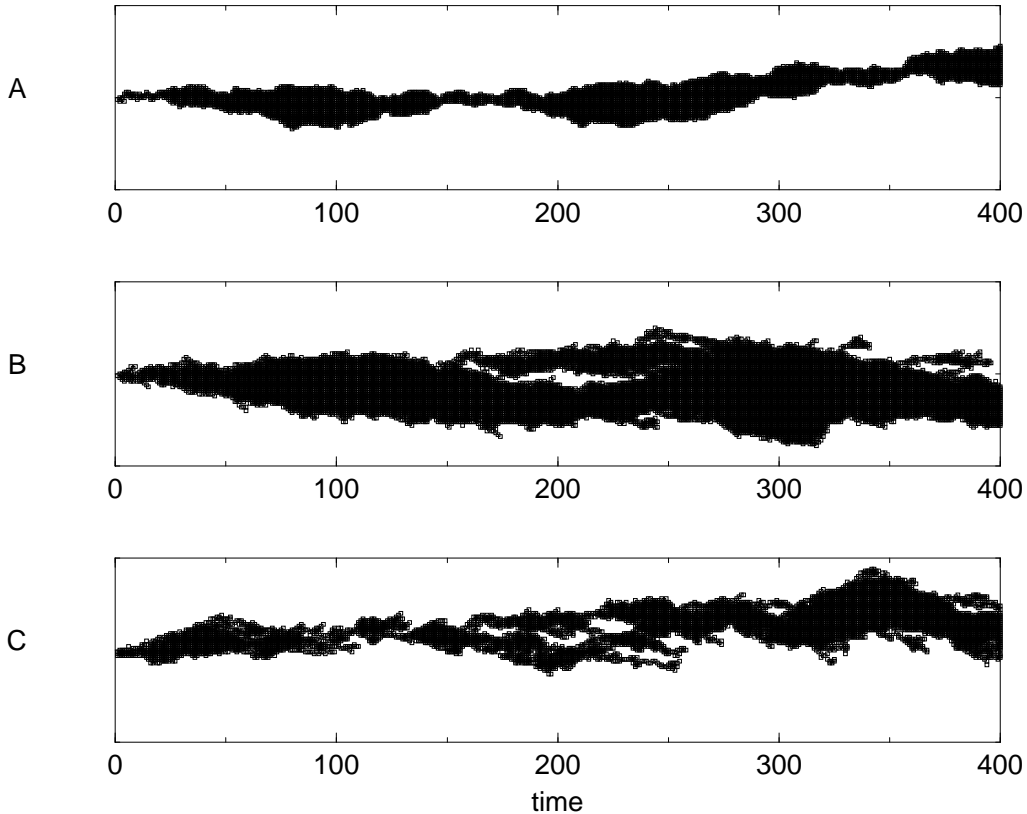}
                    \vspace*{4mm}
  }
\caption{Clusters developing from a single up-spin(dark point) in the
sea of down-spins(white points) at $t=0$ for three choices of NEKIM
parameters, see text}
\label{fig8}
\end{figure}

The hyperscaling law for the spreading exponents was derived
in its most general form by Mendes et al. \cite{mend} which we
 write here
 for the spin-quantities:
\begin{equation}
(1+\frac{\beta_s}{\beta_s{'}})\delta_s+\eta_s=dz_s/2
\label{eq:mendes}
\end {equation}
In eq.(\ref{eq:mendes}) , in analogy with the spin cluster
growth description at and in the vicinity of the CDP transition
of the DKCA \cite{DiTr},
  the above finite value of $\beta_s{'}$ enters.
(As explained in the introduction, $(p_1-p_{1c})$ of the DKCA, 
with $p_{1c}=1/2$,
 corresponds to
$h$ in the Glauber-Ising formulation). Moreover, $\beta_s=0$,
which  value follows near the PC point from the same symmetry 
consideration
as at the Glauber-Ising point (though does not in the active phase). 
Here, again, one should recall the above mentioned analogy between
the DKCA's variable $(p_1-1/2)$ and the variable $h$ in the present
case.
With the exponents obtained and summarised on Table 1 eq.(\ref{eq:mendes})
is fulfilled.
As already mentioned in the Introduction, according to the argumentation 
of \cite{DiTr} the fulfillment of the
hyperscaling law in the above form is equivalent to
compactness of the clusters. For illustration  developing clusters 
are exhibited on Fig.\ref{fig8} under three conditions:
A) Glauber case (CDP in the DKCA sense), B) at the NGI point (see Fig.\ref{fig3})
where the kinetics is  a non-equilibrium one ($p_{ex}\neq 0$) and
C) at the PC point. It is apparent that the minority phase never
develops inside of the majority one, moreover, 
the branching process present in the kinetics  in cases B) and C) 
makes  the fla pieces of CDP  boundaries fringed.

The results together with some of the critical exponents obtained earlier in
\cite{MEOD96} are summarized in Table I.
\begin{table}[h]
\begin{tabular}{|l|l|l|l|l|l|l|l|}
     & $\beta_s$&$\nu$& ${\beta_s}'$&$\Delta$&$\eta_s$&$\delta_s$&$z_s$ \\
\hline
NGI-CDP &$0$&$1/2$& $.99(2)$ & $1/2$ & $.0006(4)$ & $.500(5)$&$1(=2/Z)$ \\
\hline
CPC &$ .00(2)$&$.444$ &$.45(1)$&$.49(1)$ & .288(4) & $.287(3)$&$1.14(=2/Z)$ \\
\end{tabular}
\caption{\em Spin-cluster  critical exponents
for NEKIM in a magnetic field}
\end{table}

\section{summary}
In summary, we have carried out numerical studies of the power law
behaviour of spreading of spins,
 at the PC transition
 point of NEKIM (where a second order transition occurs on the
 level of {\it kinks}). It has been found that the analogue of the 
 Domany-Kinzel CDP transition 
 - a first order transition upon
 changing the sign of
 an applied magnetic field - still exists.
 Of the three exponents measured only 
 $\Delta$, 
 which is the static 
 magnetic exponent of the Ising model,was found to be unaffected by
 the critical fluctuations of the kinks within error. Concerning other static 
 Ising exponents,
 this circumstance
 is not so natural, as e.g. the coherence length exponent is a counterexample
 (see ref.\cite{MEOD96} and Table 1). Consequently the relation 
 $\Delta=\nu$ valid
 in the Glauber-Ising case is no more fulfilled at the PC point.
 $\beta_s{'}$ characterizing the level-off values
 of the survival probability of the spin clusters is a new (static) exponent;
  $\Delta$ and $\beta_s{'}$ are connected by a
 scaling law. The third exponent, $\eta_s$ has proven to be numerically 
 equal to
 $\delta_s=\delta$ thus ensuring that the hyperscaling law
 is fulfilled in a form appropriate for first order transitions
 and compact clusters \cite{mend,DiTr}. Moreover,we have reported 
results of  simulation for exponent $\delta$ in case of NEKIM
for the first time.

Our present results give further evidence 
to the conclusion that the effect of fluctuations felt
by the spin system at the PC transition is of interest in itself.

{\bf ACKNOWLEDGEMENTS}\\
The authors thank Z.R\'acz for useful remarks.Support from 
  the Hungarian research fund OTKA ( Nos.
T017493, 027391 and 023552) and from NATO grant CRG-970332 
is gratefully acknowledged.
 The simulations were partially carried out on
the Fujitsu AP1000 parallel supercomputer.

\end{document}